\documentclass[12pt,twocolumn]{iopart}
\usepackage{iopams}
\usepackage{graphics}

\begin{document}

\title[Quantum hologram of macroscopically entangled light]{Quantum hologram of macroscopically entangled
light via the mechanism of diffuse light storage}

\author{L.V. Gerasimov${}^1$, I.M. Sokolov${}^1$, D.V. Kupriyanov${}^1$,\\  M.D. Havey${}^2$}

\address{${}^1$Department of Theoretical Physics, St-Petersburg State
Polytechnic University, 195251, St.-Petersburg, Russia\\
${}^2$ Department of Physics, Old Dominion University, Norfolk, VA 23529}
\ead{kupr@dk11578.spb.edu}
\begin{abstract}
In the present paper we consider a quantum memory scheme for light diffusely
propagating through a spatially disordered atomic gas. The diffuse trapping of the signal
light pulse can be naturally integrated with the mechanism of stimulated
Raman conversion into a long-lived spin coherence. Then the quantum state of
the light can be mapped onto the disordered atomic spin subsystem and can be stored in
it for a relatively long time. The proposed memory scheme can be applicable for
storage of the macroscopic analog of the $\Psi^{(-)}$ Bell state and the prepared
entangled atomic state performs its quantum hologram, which suggests the possibility of further
quantum information processing.
\end{abstract}


\pacs{34.50.Rk, 34.80.Qb, 42.50.Ct, 03.67.Mn}

%% 34.50.Rk, 34.80.Qb - Laser modified scattering and reactions
%% 42.50.Ct - Quantum description of interaction of light and
%%            matter, related experiments
%% 03.67.Mn - Entanglement measures, witnesses, and other characterizations

\vspace{2pc} \noindent{\it Keywords}: Cold atoms, Light storage and
quantum memory, Entangled states
%\submitto{\JOB}
\maketitle

\section{Introduction}

At present cold atomic systems have shown themselves as promising
candidates for effective light storage, see recent reviews of the problem in
Refs.\cite{PSH,Simon}. However further improvement of atomic memory
efficiencies in either cold or warm atomic vapors is a rather challenging, and not so straightforward, experimental
task. In the case of warm atomic vapors any increase of the sample optical depth
meets a serious barrier for the electromagnetically induced transparency (EIT)
effect because of the rather complicated, and mostly negative, influence of atomic motion
and Doppler broadening, which manifest themselves in destructive interference
among the different hyperfine transitions of alkali-metal atoms
\cite{MSLOFSBKLG}. In the case of cold and dilute atomic gases, prepared for instance in a
magneto-optical trap (MOT), for some unique experimental designs an optical
depth around several hundred is attainable \cite{FGCK}, but there is are certain challenges
in accumulation of such a large number of atoms in a MOT and in making such a
system controllable.

One possible solution implies special arrangements of effective light storage
in a cold atomic sample in the diffusive regime, see \cite{GSOH}. In this case, in an
optically dense atomic sample with a given number of atoms, the actual random
optical path of light transport becomes much longer than for the single passage
of the same sample in the forward direction either under conditions of the EIT
effect or in the near resonance transparency spectral window. As a rough
estimate, if for an atomic medium formed in a MOT the optical depth on a closed
resonance transition is $b_0$, then the actual diffusive path can be
$b_{\Sigma}\sim b_0^2$ i. e. $b_0$ times longer. For typical parameters
$b_0\sim$ $20 - 50$ we have a very promising enhancement resource for the light
storage via stimulated Raman conversion of the signal pulse as it interacts
with the atomic sample in the diffusive regime.

In the present report we discuss such a diffusive quantum memory mechanism in
the context of its application to storage of macroscopically entangled light.
One experimental technique for generation of polarization entangled light by
near subthreshold SPDC type-II light source is well established now, see
Ref.\cite{ICRL}. An alternative approach under development utilizes the
polarization self rotation effect to generate significantly polarization
squeezed light Refs.\cite{Matsko}-\cite{Barreiro}. This approach has the
substantial advantage of generating quantum states of light with narrow
bandwidth and tunability in the vicinity of atomic resonances, important
characteristics for an atomic physics based quantum memory.  In either case,
the generated light possesses the quantum information encoded by some mechanism
and, in the case of Ref. \cite{ICRL}, in the strongly correlated photon numbers
related to different polarization modes. Each mode can be stored in a
particular memory unit via transforming the unknown number of photons to the
equally unknown number of atoms repopulating the signal level. Importantly, the
specifics of the considered memory scheme and of the stored quantum state is
that there is no need for further recovering of the signal light in its
original mode or modes. We show that all the stored quantum correlations can be
observed in the standard interferometric technique via relevant operations with
the prepared quantum hologram.

\section{The mechanism of diffuse light storage}
\begin{figure}[t]
\centerline{\includegraphics{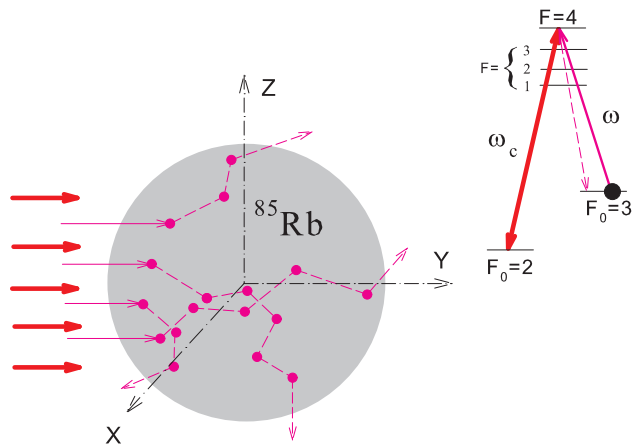}} \caption{(Color online)
The mechanism of diffuse storage of light for the example of light trapping on the $F_0=3\to F=4$ closed transition
in ${}^{85}$Rb. The diffuse propagation of a signal mode of frequency $\omega$ is indicated by
pink thin lines and arrows. The diffusion process is affected by a strong control
mode of frequency $\omega_c$ indicated by red and thick arrows. This converts a signal pulse into
a long-lived spin coherence in the atomic subsystem.}
\label{fig1}
\end{figure}

The considered light storage mechanism is based on stimulated Raman conversion
of a signal pulse into a long-lived spin coherence. The main difference with
traditional approaches, see Refs.\cite{PSH,Simon}, requires that the pulse
diffusely propagates through an atomic sample. The hyperfine energy structure
of heavy alkali-metal atoms, such as rubidium or cesium, allows convenient
integration of these processes Ref.\cite{GSOH}. In Figure \ref{fig1} we
illustrate this through the example of the $F_0=3\to F=4$ closed transition in
${}^{85}$Rb. The crucial point for the protocol is the presence and strong
action of the control laser mode repopulating the atoms from the background
$F_0=3$ hyperfine sublevel to the signal $F_0=2$ sublevel via Raman interaction
with the upper $F=3,2$ states. As a result, the quantum state of light can be
mapped onto a disordered hyperfine coherence and collectivized by the atomic
ensemble. Under ideal conditions, without relaxation and atomic losses, the
stored state effectively performs a quantum hologram of the signal light, which
can survive a relatively long time in the atomic spin subsystem.

To make such a memory scheme feasible, the following two important criteria
should be fulfilled. In the Raman mechanism only those pulses can be
effectively delayed, which have relatively narrow spectral width
\begin{equation}
\Delta\omega_{\mathrm{pulse}}\ll \Gamma_{\mathrm{AT}}b_{\Sigma}\sim%
O(1)\,\overline{\frac{\Omega_c^2}{\Delta^2}}\gamma\, b_{\Sigma}%
\label{1}%
\end{equation}
Here $\Gamma_{\mathrm{AT}}\sim
O(1)\overline{\frac{\Omega_c^2}{\Delta^2}}\gamma$, where $\gamma$ is the
natural radiative decay rate, is an estimate for the bandwidth of the
Autler-Townes (AT) resonance, created by the control field. The bandwidth is
expressed by the averaged Rabi frequency for the control field $\Omega_c$ and
by its averaged detuning $\Delta$ from those upper state hyperfine sublevels,
which are involved in the Raman process. In our case $\Delta$ can be estimated
by the hyperfine splitting $\Delta_{\mathrm{hpf}}$ between the $F=4$ and $F=3$
sublevels in the upper state. The dimensionless optical depth $b_{\Sigma}$ can
be estimated as $b_{\Sigma}\sim n_0(\lambda/2\pi)^2L_{\Sigma}$, where $n_0$ is
a typical density of atoms in the sample, $\lambda$ is the radiation wavelength
and $L_\Sigma$ is the length of a diffusive path of the signal pulse in the
sample, see Figure \ref{fig1}. Physically the condition (\ref{1}) constrains
the spectral domain where dispersion effects are manifestable. However for the
spectrally narrow pulses there can be a strong influence of the spontaneous
Raman losses initiated by the scattering on the absorption part of the AT
resonance. To guarantee that the spontaneous scattering is a negligible effect,
the following inequality, as an alternative to (\ref{1}), should be fulfilled
\begin{equation}
\Delta\omega_{\mathrm{pulse}}\gg \Gamma_{\mathrm{AT}}\sqrt{b_{\Sigma}}\sim%
O(1)\,\overline{\frac{\Omega_c^2}{\Delta^2}}\gamma\,\sqrt{b_{\Sigma}}%
\label{2}%
\end{equation}
Considered together both the inequalities (\ref{1}) and (\ref{2}) leads to an
analog of the well known requirement for atomic memory units $b_{\Sigma}\gg 1$,
see Ref.\cite{PSH}, i. e. in our case the signal pulse should pass a long
transport path in the medium. As we pointed out above, in the diffusive regime
this path can be made very long.

\begin{figure}[t]
\centerline{\includegraphics{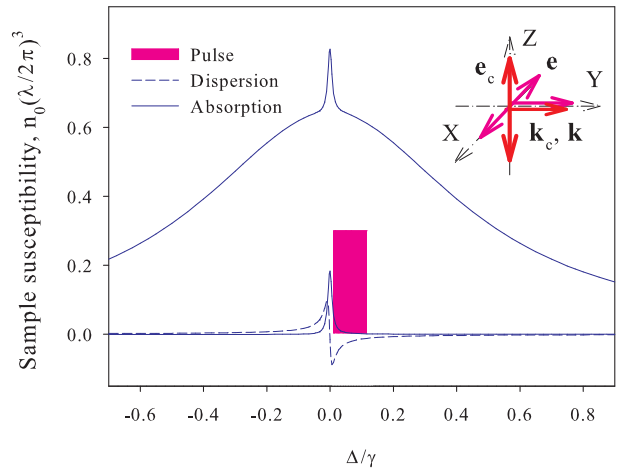}} \caption{(Color online) Dielectric susceptibility of
the sample versus the pulse spectrum. The spectra are shown in the vicinity of the $F_0=3\to F=4$ resonance line
of ${}^{85}$Rb and $\Delta$ is the relevant frequency detuning. The susceptibility is scaled
by the dimensionless density of atoms $n_0(\lambda/2\pi)^3$. The upper curve indicates the overall
absorption profile modified by the presence of the control mode and the lower curves select
the contribution of the absorption and dispersion parts of the Autler-Townes resonance only. The initial
polarization and propagation directions for the signal mode $\mathbf{e},\mathbf{k}$ as well as the
polarization and propagation directions of the control mode $\mathbf{e}_c,\mathbf{k}_c$ are
related to the reference frame and the excitation geometry shown in Figure \ref{fig1}.
The pulse spectrum is shown by the pink filled area.}
\label{fig2}
\end{figure}

The optimal pulse spectrum seems $\Delta\omega_{\mathrm{pulse}}\lesssim \gamma$
which justifies the substantial trapping of light via the multiple scattering
mechanism. In this case the variation of the Rabi frequency of the control mode
is bounded by the above inequalities, which together comprise the losses and
dispersion effects. In reality even a standard one dimensional realization of
the quantum memory protocol requires serious optimization efforts, see
Ref.\cite{PSH}. Apparently in the discussed three dimensional configuration,
which is principally based on the $D_2$-line multilevel energy structure of
alkali-metal atom, the optimization scheme is expected to be much more
complicated. Inequalities (\ref{1}) and (\ref{2}) give us a physically clear
but only rough approximation, which we can consider as only the simplest
qualitative recommendation. In Figure \ref{fig2} a typical spectral dependence
for the dielectric susceptibility of the atomic sample, modified by the
presence of the control mode, is shown. The spectra are reproduced in the
vicinity of the $F_0=3\to F=4$ resonance line of ${}^{85}$Rb and $\Delta$ is
the relevant frequency detuning. The sample susceptibility is scaled by the
dimensionless density of atoms $n_0(\lambda/2\pi)^3$ and we address the reader
to Ref.\cite{GSOH} to see the calculation details. The upper curve indicates
the overall "absorption" profile, which is actually responsible not for
absorption but for incoherent light scattering, and the lower curves select the
contribution of the absorption and dispersion parts of the AT-resonance only.
The initial polarization direction for the signal mode $\mathbf{e}$ and the
polarization of the control mode $\mathbf{e}_c$ are related to the reference
frame and the excitation geometry shown in Figure \ref{fig1}. The spectrum of
the signal pulse, shown by the pink filled area, is narrower than the original
Lorentz profile of the non-disturbed atomic resonance $F_0=3\to F=4$ but
essentially broader than the AT resonance. We consider the flat spectral
profile to follow how the photons having a frequency uncertainty randomly
distributed in the selected spectral area could be potentially delayed via the
diffuse memory protocol.

\begin{figure}[t]
\centerline{\includegraphics{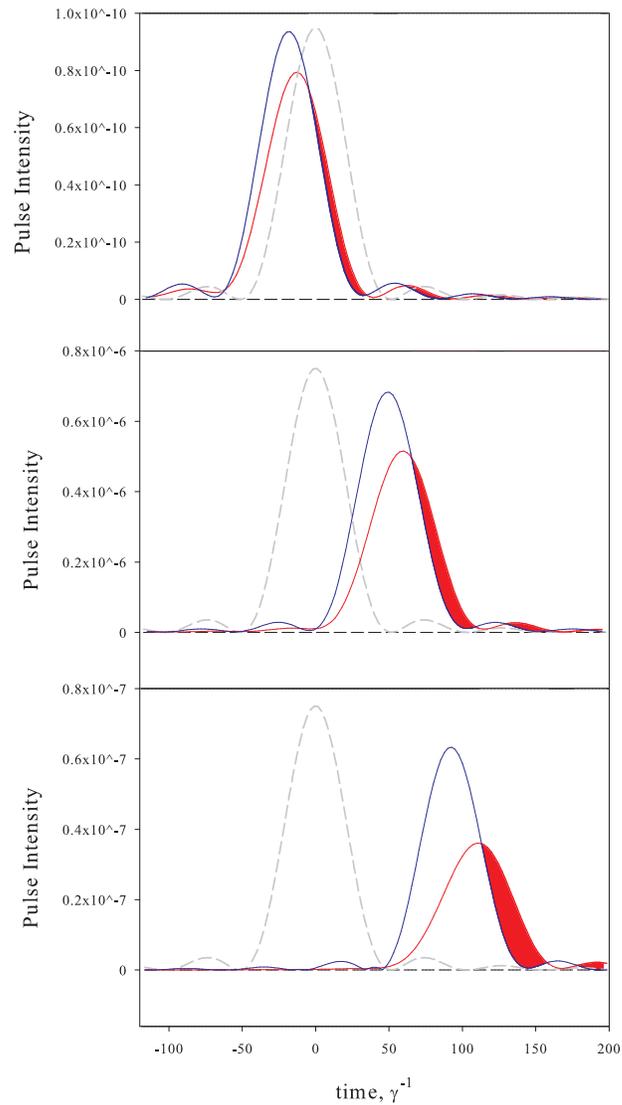}} \caption{(Color online) These graphs
subsequently show the delay induced by the control field to
those pulse fragments, which are freely passed forward (upper panel), and scattered in
the forward direction in the
50-th (central panel) and the 100-th (lower panel) orders of multiple scattering.
The gray dashed curve indicates the original profile (with arbitrary scaled
amplitude) of the pulse incident on the sample }
\label{fig3}
\end{figure}

In Figure \ref{fig3} we demonstrate a portion of our Monte-Carlo simulations
of the process. The performed calculations have been done for a spherical
atomic cloud consisting of ${}^{85}$Rb atoms with a Gaussian-type radial
distribution characterized by a squared variance $r_0^2$. The optical depth for a
light ray propagating through the central point of the cloud is given by
$b_0=\sqrt{2\pi}\,n_0\,\sigma_0\,r_0$, where $n_0$ is the peak density of atoms
and $\sigma_0$ is the resonance cross-section for the $F_0=3\to F=4$
transition. In our calculations we used  $b_0=20$, which is an example of
attainable depth in cold atom experiments with alkali-metal atom samples prepared in a MOT. The
graphs of Figure \ref{fig3} subsequently show how the delay effect associated
with the control field is accumulated as the scattering order is increased. In
our numerical simulations we assumed the simplest atomic distribution with
equal population of all the Zeeman sublevels in the background state ($F_0=3$).
This creates the AT resonance with rather small amplitude, see Figure
\ref{fig2}. The resonance could be essentially enhanced for the atomic ensemble
consisted of the spin oriented atoms, which would select the $\Lambda$-type
optical transitions with the highest coupling strength. However even in the
case of weak AT resonance for high scattering orders the delay effect becomes
quite visible such that a significant part of the light can be stored in the
spin subsystem. As was pointed out in the introduction, for diffusive
propagation the light can experience several hundred scattering events before
it leaves the sample.

The Monte Carlo simulations normally give a rather realistic approximation of
light diffusion to experimental situation but it cannot demonstrate the real
potential for the scheme of diffuse light storage in its optimal regime. As
known from many discussions of more simple and traditional one dimensional
realizations of either EIT or Raman memory protocols, see review \cite{PSH} and
reference therein, the "write-in" and "readout" stages of the protocol are not
completely symmetric parts of the entire process. The retrieval of the signal
pulse can be made much more effective by applying the control mode in the
backward direction, a point that is physically supported by time reversal
arguments. In particular, for the case of a stimulated Raman process in its
optimal configuration the stored signal light can be effectively mapped onto
the spin coherence localized near the edge of the atomic sample where light
enters the sample. This yields a very effective retrieval in the backward
direction, as was confirmed by the round of calculations presented in
Ref.\cite{SGSKMGL} for cesium atoms. In the discussed case we can also expect
the optimal light conversion to the spin coherence near that boundary of the
atomic cloud where the incident light penetrates. Then for the best strategy in
the write-in stage of the memory protocol, the optimal spectrum for the signal
light should be concentrated near the steep part of the dispersion such that
its further retrieval in the forward direction would be ineffective and
accompanied by strong spontaneous losses. The time reversal retrieval scheme
would be not so easy to organize in the three dimensional disordered
configuration, see Ref.\cite{GSOH}, and one can use an extremely weak control
mode in the read out stage of the protocol to solve the problem and minimize
spontaneous losses. In the present report we shall discuss one specific
application example where it is not necessary to retrieve the pulse in its
original mode and it is only important to provide its effective storage, such
that the protocol does not require the light transport throughout the whole
sample. For this special situation the above memory scheme is expected to be
much more effective than its one dimensional counterpart.

\section{Quantum hologram}

The proposed memory scheme is applicable and adjusted to the situation when the
quantum information is originally encoded in the total number of photons in the
signal light beam(s), this number considered as a quantum variable. Such variables are
insensitive to either spatial or temporal mode structure of the signal light
pulse. Physically this means that an unknown number of informative photons can
be mapped onto the atomic subsystem via Raman-induced repopulation of the
equivalent unknown number of the atoms to the signal level while light
diffusely propagates through the sample. Such a situation takes place with
storage of the macroscopic analog of the Bell state, this consisting of pairs
of photons with either (orthogonal) horizontal (H) or vertical (V)
polarizations having unknown but strongly correlated photon numbers, see Ref.
\cite{ICRL}. In the present report we shall consider, as an example, the following
entangled quantum state of light
\begin{equation}
|\Psi^{(-)}\rangle=\sum_{m,n}\Lambda_{mn}^{(-)}|m\rangle_{H1}|n\rangle_{V1}%
|m\rangle_{V2}|n\rangle_{H2}%
\label{4}%
\end{equation}
where
\begin{equation}
\Lambda_{mn}^{(-)}=(-)^{n}\frac{\bar{n}^{\frac{m+n}{2}}}{[1+\bar{n}]^{\frac{m+n}{2}+1}}%
\label{5}%
\end{equation}
which possesses completely anti-correlated polarizations in the light beams 1 and
2, these beams propagating in different directions. The unique property of this state is
that detection of a certain number of photons in any polarization (not only in
$H$ or $V$ but also in any elliptical polarization state) in beam 1 guarantees the detection of
the same number of photons in beam 2 but always in an orthogonal polarization state. In
Ref.\cite{ICRL} such a state was called a macroscopic analog of a singlet-type
two-particle Bell state. Its Schmidt decomposition, given by Eqs. (\ref{4}) and
(\ref{5}), can be found via basic expansion for the two mode squeezed state, see
Ref. \cite{Braunstein}. The quantum state (\ref{4}) can be parameterized by the
average number of photons in each light beam $\bar{n}$.

\begin{figure}[t]
\centerline{\includegraphics{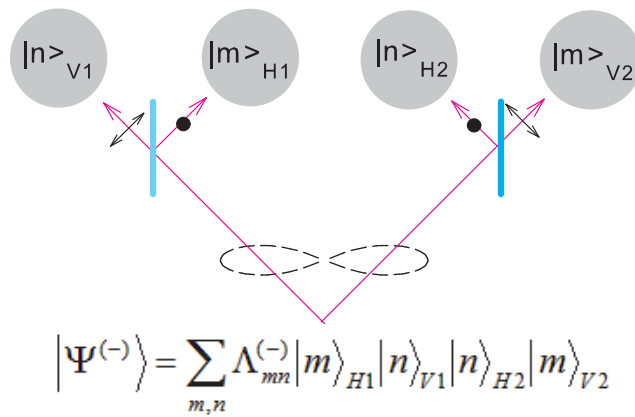}} \caption{(Color online) The quantum hologram
of the macroscopically entangled state of light $|\Psi^{(-)}\rangle$, which can be prepared
by a SPDC process, see Ref.\cite{ICRL}. The unknown photon numbers in each polarization are
subsequently stored in four memory units.}
\label{fig4}
\end{figure}

In Figure \ref{fig4} we illustrate how the quantum hologram of this light can
be realized with the memory units, as described above. Each memory unit stores
the photons of a particular polarization and frequency, which diffusely
propagate through an atomic cloud and are substantially trapped on the closed
transition and the stimulated Raman process is initiated by interaction with
the control mode  $\omega_c$ on other hyperfine sublevels ($F=3, 2$) as is
shown in the transition diagram of Figure \ref{fig1}. The more subtle point of
this process is that it creates a pure quantum state in the atomic subsystem,
which is entangled among the four clouds and has an unknown but strongly
correlated number of atoms repopulated onto the signal level $F_0=2$ in each
cloud, such that their further measurement would demonstrate the presence of
quantum non-locality in the matter subsystem.

A standard strategy for a quantum memory normally aims towards a goal of
recovery of the signal pulse in its original mode. In our situation there is no
need to do that since all the quantum information is encoded into the numbers
of repopulated atoms. The observation or detection of these numbers can be
organized with the Mach-Zehnder interferometer, as is shown in Figure
\ref{fig4}. The interferometer can be adjusted for balanced detection of the
signal expressed by the difference of the photocurrents from the output ports.
Then the measured signal associated with the small informative phase shift
$\delta\phi$, induced in one arm of interferometer, is given by
\begin{equation}
i_{-}=i_{1}-i_{2}\propto \bar{i}\,\delta\phi,\ \ \ \delta\phi=\xi\, n%
\label{6}%
\end{equation}
where the phase shift is proportional to the number of detected atoms $n$ and
to a small factor $\xi\ll 1$, which depends on geometry (sample size, aperture
of the light beam etc.) and reflects the weakness of the signal. The $D_2$-line
energy structure of an alkali-metal atom allows one to tune the probe near the
resonance associated with the closed transition and to avoid any negative
presence of the Raman scattering channel. In the case of ${}^{85}$Rb, that is
the $F_0=2\to F=1$ transition. Then the standard sensitivity of the measurement
is limited by the shot-noise, Poissonian level, but it can be essentially
improved via sending a portion of squeezed light to the second port of
interferometer, see Ref.\cite{Kimble}.

\begin{figure}[t]
\centerline{\includegraphics{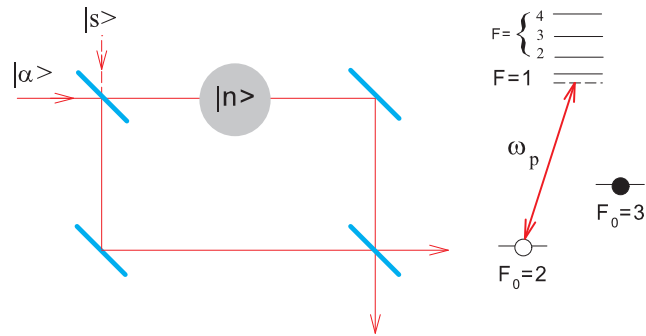}} \caption{(Color online) Schematic diagram
of the Mach-Zehnder interferometer for detecting a small number of atoms stored in a particular cloud.
The weak probe coherent mode $|\alpha\rangle$ of frequency $\omega_p$ is applied near
the resonance of the closed $F_0=2\to F=1$ transition
to avoid effects of Raman scattering. The sensitivity of the interferometer can be enhanced via sending
a portion of the squeezed light $|s\rangle$ to the second input port of the interferometer, see Ref. \cite{Kimble}.}
\label{fig5}
\end{figure}

Let us make the following remark concerning the above detection scheme, which
in an ideal situation would perform a certain type of quantum non demolition
measurement (QND), see Ref.\cite{WolfMandel}. At first sight the scheme seems
it is specifically related to the polarization basis $|H\rangle$ and
$|V\rangle$, which was used in expansion (\ref{4}) and in the storage protocol
shown in Figure \ref{fig4}. However, an identical expansion could be rewritten
in any other basis of arbitrary orthogonal elliptical polarizations and this
would be described by the same expansion coefficients. In other words the state
$|\Psi^{(-)}\rangle$ is insensitive to the type of polarization beamsplitters
used for the hologram creation. Then the above QND operation with the hologram
can be interpreted as postponed detection of the photons transmitted by the
particular beamsplitters. With variation of the beamsplitter types the
measurement statistics could demonstrate violation of the classical probability
principles. In a particular case of a rare flux consisting of the photons'
pairs prepared in a "singlet state" the measurements would show violation of
the Bell inequalities. We can also point out that the hologram yields various
of interferometric operations and could potentially be interesting as a logic
element for the further quantum information processing based on a continuous
variables scheme.

\section*{Acknowledgements}
We thank Maria Chekhova and Timur Iskhakov for fruitful discussions, which
initiated this work. The work was supported by RFBR (grant 10-02-00103) and NSF
(grant NSF-PHY-1068159) and by Federal Program "Scientific and
scientific-pedagogical personnel of innovative Russia on 2009-2013" (contract
\#14.740.11.0891).

\end{document}